\documentclass[twocolumn,
groupedaddress,
superscriptaddress,
amsmath,amssymb,
aps,
pra,
]{revtex4-1}
\usepackage{amsmath}
\usepackage{subfigure}
\usepackage{graphicx}% Include figure files
\usepackage{hyperref}
\hypersetup{
	colorlinks = true, linkcolor = blue, anchorcolor = blue, urlcolor = blue, citecolor = blue
}
\usepackage{xcolor}
\usepackage{verbatim}

\begin{document}
	
	%\preprint{}
	
	\title{Securing quantum key distribution systems using fewer states}
	
	
	%\email{nti3@duke.edu}
	\author{Nurul T. Islam}
	\email[]{nurul.islam@duke.edu}
	%\thanks{}
	\affiliation{Department of Physics and the Fitzpatrick Institute for Photonics, Duke University, Durham, North Carolina 27708, USA}
	\author{Charles Ci Wen Lim}
	\email[]{charles.lim@nus.edu.sg}
	\affiliation{Department of Electrical and Computer Engineering, National University of Singapore, 117583, Singapore}
    \affiliation{Centre for Quantum Technologies, National University of Singapore, 117543, Singapore}
	\author{Clinton Cahall}
	\affiliation{Department of Electrical Engineering and the Fitzpatrick Institute for Photonics, Duke University, Durham, North Carolina 27708, USA}
	\author{Jungsang Kim}
	\affiliation{Department of Electrical Engineering and the Fitzpatrick Institute for Photonics, Duke University, Durham, North Carolina 27708, USA}
    \affiliation{IonQ, Inc., College Park, MD 20740, USA}
	\author{Daniel J. Gauthier}
	\affiliation{Department of Physics, The Ohio State University, 191 West Woodruff Ave., Columbus, Ohio 43210 USA}


\date{\today}

\begin{abstract}
Quantum key distribution (QKD) allows two remote users to establish a secret key in the presence of an eavesdropper. The users share quantum states prepared in two mutually-unbiased bases: one to generate the key while the other monitors the presence of the eavesdropper. Here, we show that a general $d$-dimension QKD system can be secured by transmitting only a subset of the monitoring states.  In particular, we find that there is no loss in the secure key rate when dropping one of the monitoring states.  Furthermore, it is possible to use only a single monitoring state if the quantum bit error rates are low enough. We apply our formalism to an experimental $d=4$ time-phase QKD system, where only one monitoring state is transmitted, and obtain a secret key rate of $17.4 \pm 2.8$~Mbits/s at a 4~dB channel loss and with a quantum bit error rate of $0.045\pm0.001$ and $0.037\pm0.001$ in time and phase bases, respectively, which is 58.4\% of the secret key rate that can be achieved with the full setup. This ratio can be increased, potentially up to 100\%, if the error rates in time and phase basis are reduced. Our results demonstrate that it is possible to substantially simplify the design of high-dimensional QKD systems, including those that use the spatial or temporal degrees-of-freedom of the photon, and still outperform qubit-based ($d = 2$) protocols.

%*** Is this true? Do we have the data to support this? *** 
\end{abstract}

%\keywords{}

\maketitle

\section{Introduction} 
Quantum key distribution (QKD) is a symmetric encryption technique that allows two remote users, called Alice and Bob, to share a secret key in the presence of an eavesdropper, known as Eve~\cite{BarnettBook,GisinRMP02,Valero09}. Eve can attack the QKD system using any resources allowed by the laws of quantum physics, including the use of a quantum computer, which contrasts with conventional encryption methods that rely on potentially vulnerable hard computation problems. 

There is currently great interest in developing QKD systems that use high-dimensional quantum states (dimension $d$) because of their higher noise tolerance and increased photon information efficiency~\cite{Brougham13,Nunn13,Shapiro14,Boyd14,HD16,Ding16,Boyd17}. Furthermore, high-dimensional protocols based on time-phase states allow for higher key rates for low-loss channels appropriate for metropolitan networks when considering the practical issue of detector saturation~\cite{Taimur2017}. In fact, all current state-of-the-art QKD systems capable of generating secret key rates exceeding 20 Mbits/s are realized using high-dimensional protocols~\cite{MIT16,Taimur2017}. Nonetheless, the increased system performance of these protocols come at the cost of increased complexity in the experimental setup, which makes some of these high-dimensional systems challenging to implement, especially for field applications.

Security in QKD systems arises from the use of two different bases by Alice and Bob that are mutually unbiased with respect to each other in the simplest approach. In greater detail, consider a prepare-and-measure scheme, where Alice prepares and sends to Bob one of the $d$ quantum states in one of two bases.  When she prepares a state in one basis, Bob will measure the received state with a high degree of accuracy if he uses the same measurement basis.  On the other hand, Bob will have a high error (1-$(1/d)$) if he performs his measurement in the mutually-unbiased basis.  In a typical QKD session, Alice uses two independent quantum random number generators to generate the key and to make a preparation-basis choice, and Bob uses a quantum random number generator to make a measurement-basis choice.  After the session, Alice and Bob share the basis choices over a public channel and keep only the data when the basis choices are the same; this is commonly known as sifting. 

In practice, for efficiency reasons, one basis is typically used to encode the secret key and the other basis is used to monitor the presence of Eve~\cite{lo2005efficient}. Some theoretical investigations have explored the possibility of sending fewer monitoring states, where instead of sending the complete set of monitoring states, only a few of them are employed. The advantage is that doing so may simplify the implementation, e.g., less randomness is required and possibly fewer optical elements are needed. In the case of the Bennett-Brassard QKD with qubits (BB84), this approach has been investigated in Refs.~\cite{Lo06,branciard2007zero}, where only one basis state $|+\rangle$ from the so-called phase basis $\{|+\rangle,|-\rangle\}$ is used to secure the qubit channel. However, the security analyses showed that sending fewer monitoring states lead to sub-optimal secret key rates compared to the original setting whereby both $|+\rangle$ and $|-\rangle$ are employed. Recently, Tamaki \emph{et al.}~\cite{Tamaki14} showed that the BB84 QKD protocol can be fully secured using only one monitoring state, but additional measurement statistics has to be included in the security analysis. More specifically, by exploiting the additional information gleaned from the mismatched basis statistics~\cite{RejectedDataPatent}, the authors proved that the resulting secret key rate is exactly the same as  the original BB84 QKD. Interestingly, the authors also showed that the simplified protocol is loss-tolerant: its security is highly resistant to loss-dependent attacks exploiting state-preparation flaws~\cite{GLLP04}. Recent experiments~\cite{LossTolerantExperimental,LossTolerantMDI,MIT2017} have demonstrated and confirmed the feasibility of this protocol using simplified transmitters and receivers.

In this article, we extend these results for $d= 2$ to arbitrary $d$ using semidefinite programming (SDP). In particular, we consider a family of two-basis, high-dimensional QKD protocols, where one basis is the discrete Fourier transform of the other, and is mutually unbiased with respect to the other. We first analyze the security of this generic protocol against arbitrary collective attacks for the case where Alice sends a complete set of states in both bases. We then show that a complete set of monitoring basis states is not necessary to guarantee security of this protocol: The protocol can be secured even when using just one monitoring-basis state to determine the presence of Eve as long as the channel noise is low enough. Our analysis also takes into account the outcomes of the events where Alice and Bob choose different basis, thereby extending Tamaki $\textit{et al.}$'s proof for $d > 2$. We note that our analysis is limited to the case when the state-preparation process is ideal; it remains to be seen whether securing higher dimensional QKD protocols with fewer monitoring states will lead to loss-tolerant QKD protocols. We then apply our findings to a recently demonstrated high-dimensional time-phase QKD experiment~\cite{Taimur2017}. We show that the experimental setup can be greatly simplified if only a small number of mutually unbiased basis states are used. Our results suggest that other current qubit- or qudit-based protocols can be upgraded to enhance the secure key generation rate with simple modifications to the experimental setup.

\section{Protocols and Security Framework} 
Consider a generic QKD protocol where Alice chooses a basis, $\mathsf{T}$ or $\mathsf{F}$, using a quantum random number generator, and prepares a photonic wavepacket to encode a high-dimensional alphabet, where the quantum states in the $\mathsf{T}$-basis ($\mathsf{F}$-basis) are used to generate the secret key (monitor the presence of Eve). The $\mathsf{T}$-basis states are denoted as $|t_n\rangle$, where $n = 0, ...,~d-1$. The $\mathsf{F}$-basis states are superposition of the $\mathsf{T}$-basis states with distinct phases determined by discrete Fourier transformation of the information-basis states and given by
\begin{align}
|f_n\rangle = \frac{1}{\sqrt{d}} \sum\limits_{m= 0}^{d-1} \exp\left(\frac{2\pi i n m}{d}\right)|t_m\rangle.~~~n = 0, ..., d-1 \label{DFT}
\end{align}

We consider here a prepare-and-measure protocol, but it is well known that such a scheme can be written in an equivalent entanglement-based description \cite{Valero09}, where Alice's choice of the bit value is determined by her measurement outcome. We therefore assume that Alice and Bob share an entangled state of the form $|\phi\rangle_{AB} = (1/\sqrt{d}) \sum_{n = 0}^{d-1} |t_n\rangle_A |t_n\rangle_B$ such that a projective measurement on the entangled state by Alice determines the state received by Bob. In addition, we assume that Eve's interaction with the shared quantum state is independent and identically distributed (i.i.d.), so that after the transmission of the signal states, the density matrix shared among Alice, Bob and Eve is $\rho_{ABE}$, and the state is $|\Psi\rangle_{ABE} = \sum_j \sqrt{\lambda_j} |\phi\rangle_{AB} |j\rangle_E$. Such an i.i.d. interaction of Eve in the quantum channel is known as a collective attack. The security proof against a collective attack can be promoted to the general attacks using known techniques, such as the de Finetti theorem, if the quantum states are permutationally invariant \cite{Renner05,Renner07,Renner09,RennerdeFinetti09}.

%We do not extend the proof against general attacks mainly because it has been done in many prior studies \cite{RennerdeFinetti09}.  

%*** The English in the previous sentence made no sense.  Check that what I have written in correct. ***

The primary challenge of our security analysis is to place an upper bound on the so-called phase error rate $e^{U}_\mathsf{F}$ so that a valid lower bound on the key can be obtained. The phase error rate is defined as the error rate observed when the entangled state is measured hypothetically in the $\mathsf{F}$-basis, but the actual measurements by Alice and Bob are performed in the information ($\mathsf{T}$) basis. Phase errors are not directly observed in the experiment; rather, they are estimated based on the observed quantum bit error rates from the experiment via classical random sampling techniques. We distinguish the phase error rate from the quantum bit error rates in $\mathsf{T}$ and $\mathsf{F}$ bases, which are denoted by $e_\mathsf{T}$ and $e_\mathsf{F}$, respectively. These are error rates that occur when Alice and Bob prepare and measure the quantum states in the same basis ($\mathsf{T}$ or $\mathsf{F}$), but detect different quantum states. 

To determine the maximum value of $e^{U}_\mathsf{F}$ in this protocol, we cast it into a maximization-SDP problem, where we use the \textit{a priori} known statistics of the compatible positive-operator valued measure (POVM) of Alice and Bob. Similar SDP-based security analyses were recently presented in Ref.~\cite{Coles16}. However, our approach is different in that we are interested in bounding the phase error rate when Alice transmits to Bob less than a complete set of monitoring-basis states. Additionally, there are some conceptual differences between our approach and the more direct approach proposed by Coles \emph{et.al}~\cite{Coles16}. In particular, our approach is focused on bounding the phase error rate while the latter is focused on bounding the conditional von Neumann entropy of $\rho_{AE}$, which roughly speaking characterizes the asymptotic secret key rate of QKD assuming one-way classical communication~\cite{Devtek05}. Our approach, on the other hand, uses a well-established argument (involving the equivalence between entanglement distillation and quantum error correction via CSS codes) to bound the entropy term using the phase error rate~\cite{Shor00}. Hence, our approach is less direct, in the sense that we use one more step to bound the secret key rate. Interestingly, as we will see below, there appears to be no difference between the two approaches in terms of the achievable secret key rates, at least for the QKD protocols considered in this work. More specifically, in the case of two mutually unbiased bases with complete states, we obtain the same key rates as those predicted by earlier theoretical findings~\cite{Valero10} for $d<=7$. 

The measurement statistics can also be extracted from the experiment. For example, all the statistics of Alice and Bob's projective measurements, $\Pi_n^\mathsf{T} = |t_n\rangle \langle t_n|$ and $\Pi_n^\mathsf{F} = |f_n\rangle \langle f_n|$ where $n = 0$ to $d-1$, on the entangled state $\rho_{AB}$ are well defined and can be extracted from the experiment. In addition, the statistics of the error operators in the $\mathsf{T}$ and $\mathsf{F}$ bases, $E_\mathsf{F}$ and $E_\mathsf{T}$, respectively, are also known. Therefore, the problem can be cast as the following optimization problem
\begin{align}
\texttt{maximize:}~\mathsf{Tr}(E_\mathsf{F} \rho_{AB}) &= e^U_{\mathsf{F}}  \\
\texttt{s.t.}, 	\mathsf{Tr} (\rho_{AB}) &= 1,  \\ 
\rho_{AB} &\geq 0,  \\
\mathsf{Tr} (E_\mathsf{T} \rho_{AB}) &= e_{\mathsf{T}},  \label{Eq:opt1}\\
\mathsf{Tr} (\Pi^\mathsf{a}_{n} \otimes \Pi^\mathsf{b}_{m}  \rho_{AB}) &= p^\mathsf{a,b}_{n,m},\\ 
\forall \{a,b\} \in \{\mathsf{T}, \mathsf{F}\}~~~~&\&~~~~n,m = 0, ..., d-1 
\end{align}

In Eq.~\ref{Eq:opt1}, the quantum bit error rate $e_{\mathsf{T}}$ is measured directly in the experiment. The probabilities $p^\mathsf{a,b}_{n,m}$ of Alice sending a state and Bob receiving a state are also known. The fact that we are allowing $\rho_{AB}$ to be arbitrary also implies that Eve can perform any arbitrary operations on the states transmitted between Alice and Bob, and hence the bound is valid for any collective attack respecting the given measurement statistics. The only unknown relevant quantity in the optimization problem is the phase error rate $e^U_{\mathsf{F}}$, which we can efficiently solve using CVX, which is a Matlab software designed for convex optimization problems~\cite{cvx}. Explicit calculation of these operators is shown in Appendix A and Appendix B. 

To demonstrate that this method of optimization validates previously known bounds~\cite{Cerf02, Valero10}, we first calculate the secret key fraction defined as the number of bits per received state, which is given by
\begin{align}
K:= \log_2d - h(e^U_\mathsf{F}) - \Delta_{\mathsf{Leak}},
\end{align}
where $h(x):= -x \log_2(x/(d-1)) - (1-x) \log_2(1-x)$ is the binary entropy, $\Delta_{\mathsf{Leak}}:=h(e_\mathsf{T})$ is the fraction of the key revealed during error correction, and we have assumed that Alice and Bob exchange an infinitely-long key for simplicity.  The secure key rate is given by $R= r K$, where $r$ is the symbol preparation rate, which may depend on $d$ for some protocols.  Below, we relax the assumption of Alice using a single-photon source and consider the case whereby Alice sends weak coherent states instead of single-photon states, which is often used in practical QKD systems. 

\subsection{Secret key fraction for two-basis $d = 4$ QKD systems}

\begin{figure}[htp]
	\begin{center}
		\includegraphics[width=\linewidth]{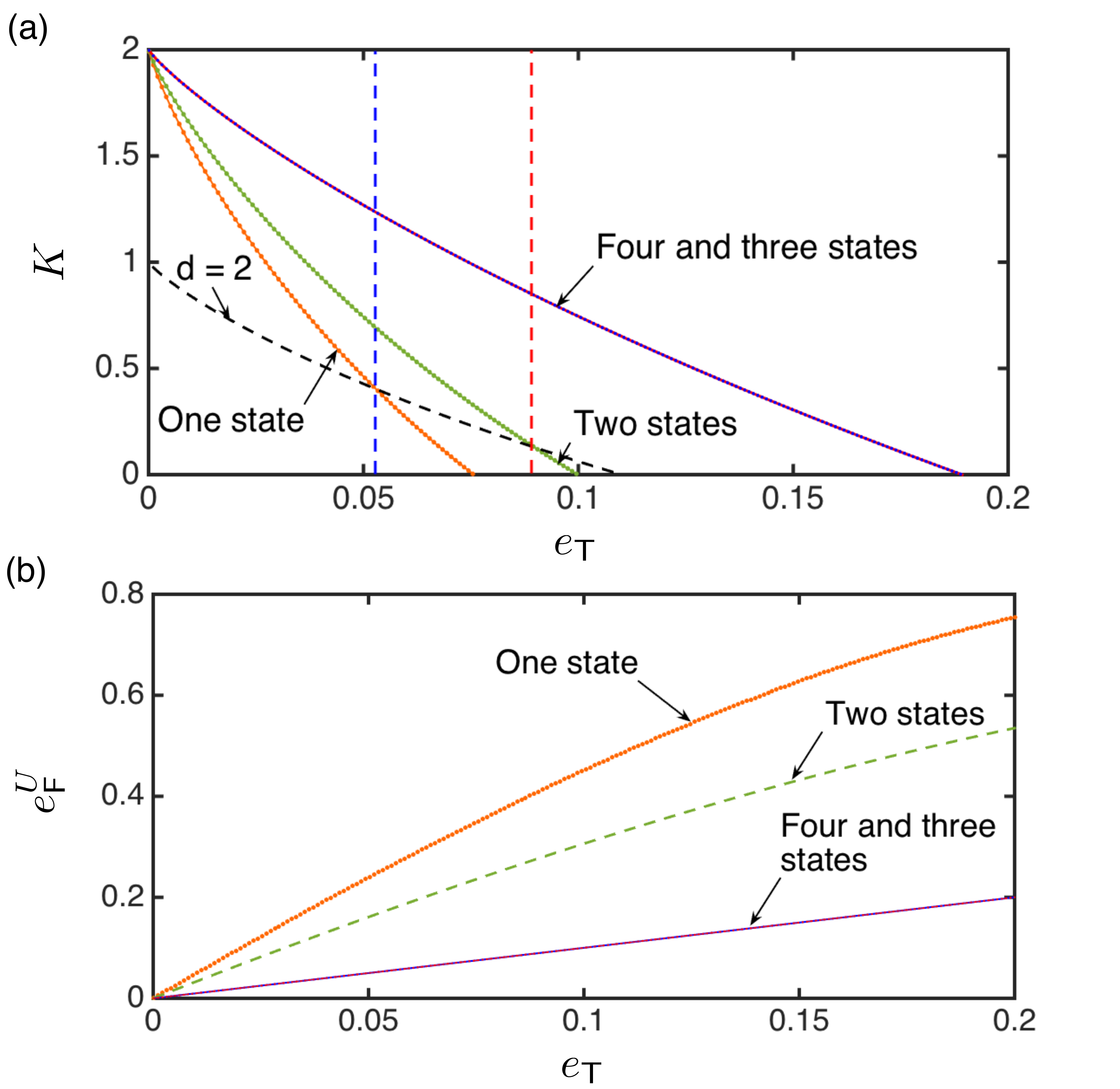}
		\caption{The secret key fraction (a) and the numerically obtained upper bound on the phase error rate (b) plotted as a function of the quantum bit error rate for $d = 4$.}
		\label{Security_Proof}
	\end{center}
\end{figure}

In Fig.~\ref{Security_Proof}(a), we show the dependence of $K$ on $e_{\mathsf{T}}$ for $d = 4$ when Alice transmits all four information-basis states and a varying number of monitoring-basis states. For reference, we also show the dependence of $K$ on $e_{\mathsf{T}}$ for $d = 2$ (black dashed line). For the specific case where Alice sends all four monitoring-basis states (solid red line), we find that the maximum error tolerance is $\sim 18.9\%$, which is in agreement with existing findings (assuming depolarizing quantum channel)~\cite{Cerf02,Valero10}. The error tolerance is defined as the error rate $e_\mathsf{T}$ beyond which $K = 0$. 

We find that $K$ is the same when Alice sends only three monitoring-basis states (Fig.~\ref{Security_Proof}(a), dotted blue line) in comparison to the case when she sends all four monitoring-basis states, illustrating that one of the states is redundant. The redundancy of mutually unbiased basis states was previously studied analytically for BB84-type ($d=2$) protocols in Ref.~\cite{Tamaki14}. However, the same approach cannot be used when Alice transmits less than $d-1$ monitoring states. Our SDP approach extends the result of Ref.~\cite{Tamaki14} and makes it possible to analyze the security for any subset of mutually unbiased basis states. For the case where Alice sends only one state or two states in the monitoring basis, the protocol still generates a positive secret key fraction, as illustrated by green and orange lines in Fig.~\ref{Security_Proof}(a), but with a lower error tolerance (7.5\% and 10\%). Despite the lower error tolerance, we observe that the secret key fraction generated with one and two monitoring basis states in $d = 4$ are higher than the secret key fraction achieved with $d = 2$ if $e_{\mathsf{T}}$ are less than $\sim 5.3$ and $\sim 8.9\%$, as indicated with the blue and red vertical dashed lines, respectively. 

For protocols in which it takes the same duration (time window) to prepare an arbitrary dimension state, this translates into higher secret key rate. Examples of such protocols include the ones where quantum states are prepared using spatial degrees of freedom, such as OAM-QKD protocols~\cite{Boyd14}. On the other hand, protocols for which a $d = 4$ state takes twice the time-window to prepare a state compared to a $d = 2$ state, such as time-bin encoding schemes~\cite{MIT16,Taimur2017}, this still translates into a higher secret key rate if the channel loss is low (photon rate is high), and the detectors are operated near the saturation regime.

In Fig.~\ref{Security_Proof}(b), we show the SDP-obtained upper bound on the phase error rate $e^U_{\mathsf{F}}$ as a function of $e_{\mathsf{T}}$. As was observed by Tamaki \textit{et al.} for $d = 2$, we also find that the error tolerance of the protocol when Alice sends only three monitoring-basis states is identical to the case where she sends all four states. This is because complete knowledge of the remaining unused monitoring state can be reconstructed from the statistics of the $d-1$ states that are used and from the statistics of the events where Alice and Bob choose different basis. However, when Alice sends only one or two monitoring-basis states, complete knowledge of the non-transmitted states cannot be reconstructed using the experimentally determined statistics. Thus, the phase error rate increases faster than the quantum bit error rate, resulting in reduced secret key fraction and lower error tolerance as shown in Fig.~\ref{Security_Proof}(b).  

\subsection{Secret key fraction for $d = 2$ to $d =7$ with $d-1$ and one monitoring-basis state}
\begin{figure}[h]
	\begin{center}
\includegraphics[width=\linewidth]{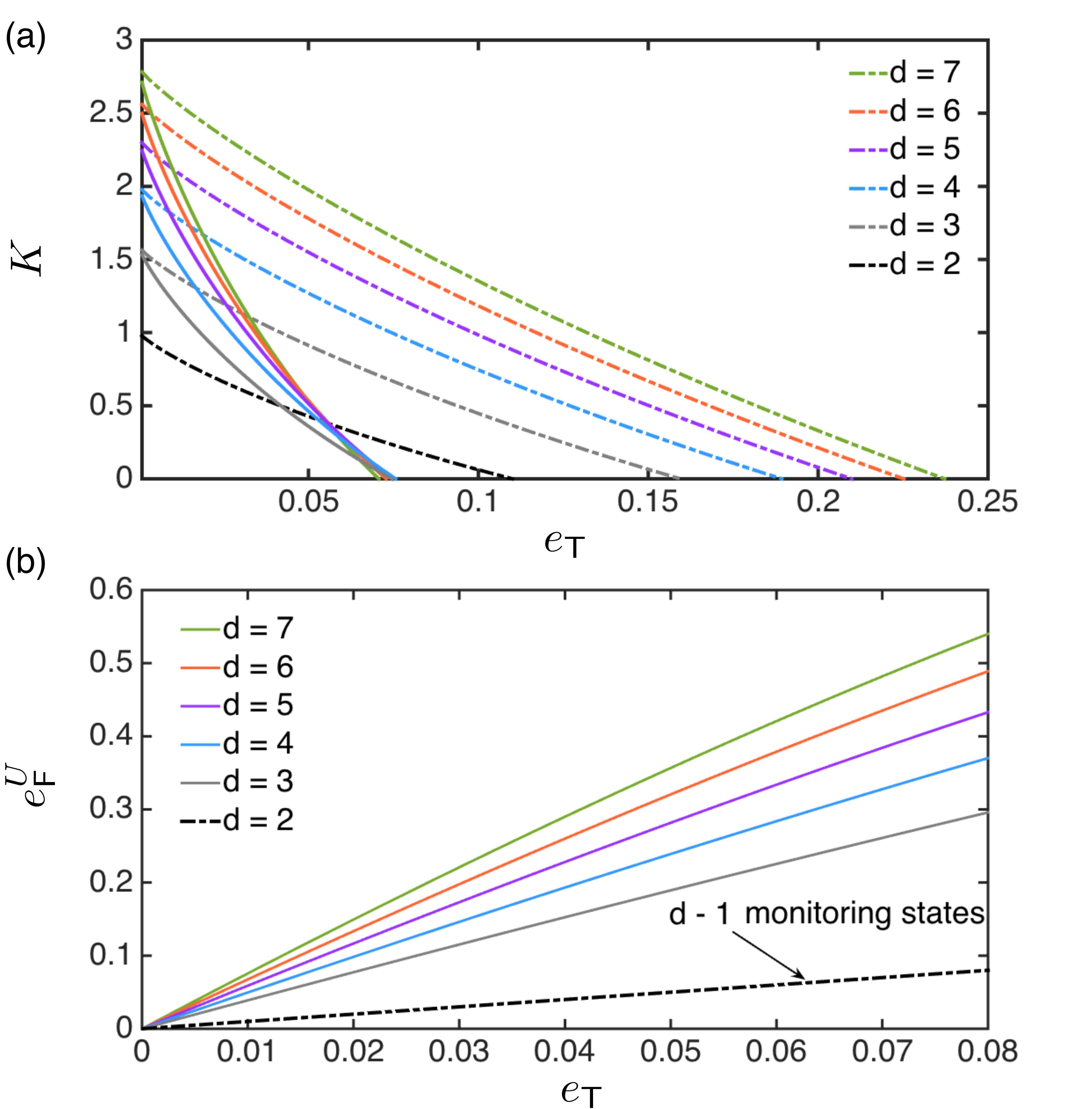}
		\caption{The secret key fraction (a) and the upper bound on the phase error rate (b) plotted as a function of the quantum bit error rate when Alice transmits $d-1$ states and only one monitoring basis state.}
		\label{Security_Proof2}
	\end{center}
\end{figure}

To demonstrate the applicability of our method for higher dimension, we consider protocols with $d$ between 2 and 7. Specifically, we obtain the secret key fraction and the upper bound on $e^U_{\mathsf{F}}$ for two specific cases: when Alice transmits $d-1$ or 1 monitoring-basis states to secure the protocol. The results are presented in Fig.~\ref{Security_Proof2}. 

In Fig.~\ref{Security_Proof2}(a), we plot the dependence of $K$ on $e_\mathsf{T}$ when Alice transmits $d-1$ monitoring basis states (dashed lines). We find that the error tolerance for all these cases are in agreement with the previously known bounds presented in Ref.~\cite{Cerf02,Valero10}. In Fig.~\ref{Security_Proof2}(b), we plot the corresponding $e_\mathsf{F}^U$ as a function of the $e_\mathsf{T}$, and observe that for all $d$ values, $e_\mathsf{F}^U = e_\mathsf{T}$, as indicated by the black dashed line. This is expected for symmetric two-basis protocol.

In Fig.~\ref{Security_Proof2}(a), we also show the dependence of $K$ on $e_\mathsf{T}$ between $d = 2~\text{and}~7$ (solid lines), when Alice transmits only one monitoring-basis state to secure the protocol. For all $d$, we observe that the higher-dimensional protocols have higher $K$ than $d = 2$ when $e_\mathsf{T}$ is relatively small. The value of $e_\mathsf{T}$ beyond which $d = 2$ will have a larger value of $K$ depends on the dimension of the system. Overall, we find that the error tolerance of all high-dimensional ($d > 2$) protocols is around 7.1-7.5\%, which is smaller than $d = 2$. However, if $e_\mathsf{T}$ is small, then the high-dimensional protocols generate more secret key per received state as shown in Fig.~\ref{Security_Proof2}(a). In Fig.~\ref{Security_Proof2}(b), we show the corresponding $e_\mathsf{F}^U$ as a function of $e_\mathsf{T}$. We find that as $d$ increases, the bound of $e_\mathsf{F}^U$ gets worse, which is in contrast with the case where Alice transmits $d-1$ states. 

In general, if the dimension of a QKD system can be changed easily, an experimentalist can estimate the expected upper bound on the phase error rate using Fig.~~\ref{Security_Proof2}(b), and then select the value of $d$ and the number of monitoring-basis states that will maximize the number of secret bits per state.

\subsection{Three-Intensity Decoy Technique}
The secret key fraction calculated above is based on an ideal single-photon source. However, most experimental implementations of QKD protocols are based on phase-randomized weak coherent sources that have photon statistics given by the Poisson distribution. It is well known that a weak coherent source with multiple decoy intensities can be used to achieve secret key rates similar to an ideal single-photon source. Here, we combine the numerics-based approach discussed above with the decoy-state technique~\cite{Lodecoy05,PracticalDecoy05} to show how to bound secret key fractions when imperfect sources are used in a QKD system.

Suppose that Alice sends quantum states with three different mean photon numbers $k \in (\mu, \nu, \omega)$, each transmitted with a probability $p_k$. The signal-state mean photon number $\mu$ is assumed to be larger than the sum of the two decoy-state mean photon numbers $\nu,~\omega$, \textit{i.e.}, $\nu + \omega < \mu$. In addition, we assume that $0\leq \omega \leq \nu$. Under these conditions, the secret key fraction can be expressed as
\begin{align} \label{SKF_Time_Phase}
K := R_{\mathsf{T},1} [\log_2 d - h\{e_\mathsf{F}^{U}(.)\}] - R_\mathsf{T}\Delta_{\mathsf{Leak}},
\end{align}
where $R_{\mathsf{T},1}$ is the single-photon gain in the $\mathsf{T}$ basis, $e_\mathsf{F}^U$ is a function of the single-photon error rate in $\mathsf{F}$ basis, denoted by $e_{\mathsf{F,1}}$, and $R_\mathsf{T}$ is the overall gain in the $\mathsf{T}$ basis. 

The single-photon gain is bounded by
\begin{align}
R_{\mathsf{T},1} = [p_\mu (\mu e^{\mu})+p_\nu (\nu e^{\nu}) + p_\omega (\omega e^{\omega})]Y_{\mathsf{T},1},
\end{align}
where
\begin{align}\label{eq:single_photon_yield}
Y_{\mathsf{T},1}  &= \max \left\{ \frac{\mu}{\mu \nu - \mu \omega - \nu^2 + \omega^2} [R_{\mathsf{T},\nu} e^{\nu} - R_{\mathsf{T},\omega} e^{\omega} \nonumber \right.\\
&\left.- \frac{\nu^2 - \omega^2}{\mu^2}(R_{\mathsf{T},\mu} e^{\mu} - Y_{\mathsf{T},0})],~0\right\}
\end{align}
is the single-photon yield, $R_{\mathsf{T},k}$ is the gain corresponding to mean photon number $k\in (\mu, \nu, \omega)$ in the $\mathsf{T}$ basis, and $Y_{\mathsf{T},0}$ is the zero-photon yield in the $\mathsf{T}$ basis bounded by
\begin{align}
Y_{\mathsf{T},0} = \max \left\{ \frac{\nu R_{\mathsf{T},\omega}e^{\omega} -
\omega R_{\mathsf{T},\nu}e^{\nu}}{\nu - \omega},~0\right\}.
\end{align}
Finally, the single-photon error rate in the $\mathsf{F}$ basis is given by
\begin{align}\label{eq:error_rate}
e_{\mathsf{F,1}} = \min \left\{ \frac{e_{\mathsf{F},\nu} R_{\mathsf{F},\nu} e^{\nu} - e_{\mathsf{F},\omega} R_{\mathsf{F},\omega} e^{\omega}}{(\nu - \omega) Y_{\mathsf{F},1}},~\frac{1}{2}\right\},
\end{align}
where $e_{\mathsf{F},k}$ is the error rate in the $\mathsf{F}$ basis with mean photon number $k$. In Eq.~\ref{eq:error_rate}, $Y_{\mathsf{F},1}$ is the single-photon yield in the $\mathsf{F}$ basis, which is obtained from Eq.~\ref{eq:single_photon_yield} by replacing $\mathsf{T}$ with $\mathsf{F}$. 

When all $d$ or $d-1$ states are transmitted in the $\mathsf{F}$ basis, $e_{\mathsf{F}}^U (.) = e_{\mathsf{F},1}$, and therefore obtaining a bound numerically is not necessary. However, when only one state in the $\mathsf{F}$ basis is transmitted to monitor the presence of an eavesdropper, $e_{\mathsf{F}}^U(.)$ can be estimated in two steps. First,  estimate the single-photon error rate in the $\mathsf{F}$ basis using Eq.~\ref{eq:error_rate}. Second, find the corresponding $e_\mathsf{F}^U$ from Fig.~\ref{Security_Proof2}(b) using the value for the single-photon quantum bit error rate found in the first step. The same procedure can be repeated for any number of monitoring states. 

\section{Experimental Demonstration} 
\begin{figure}[htp]
	\begin{center}		\includegraphics[width=\linewidth]{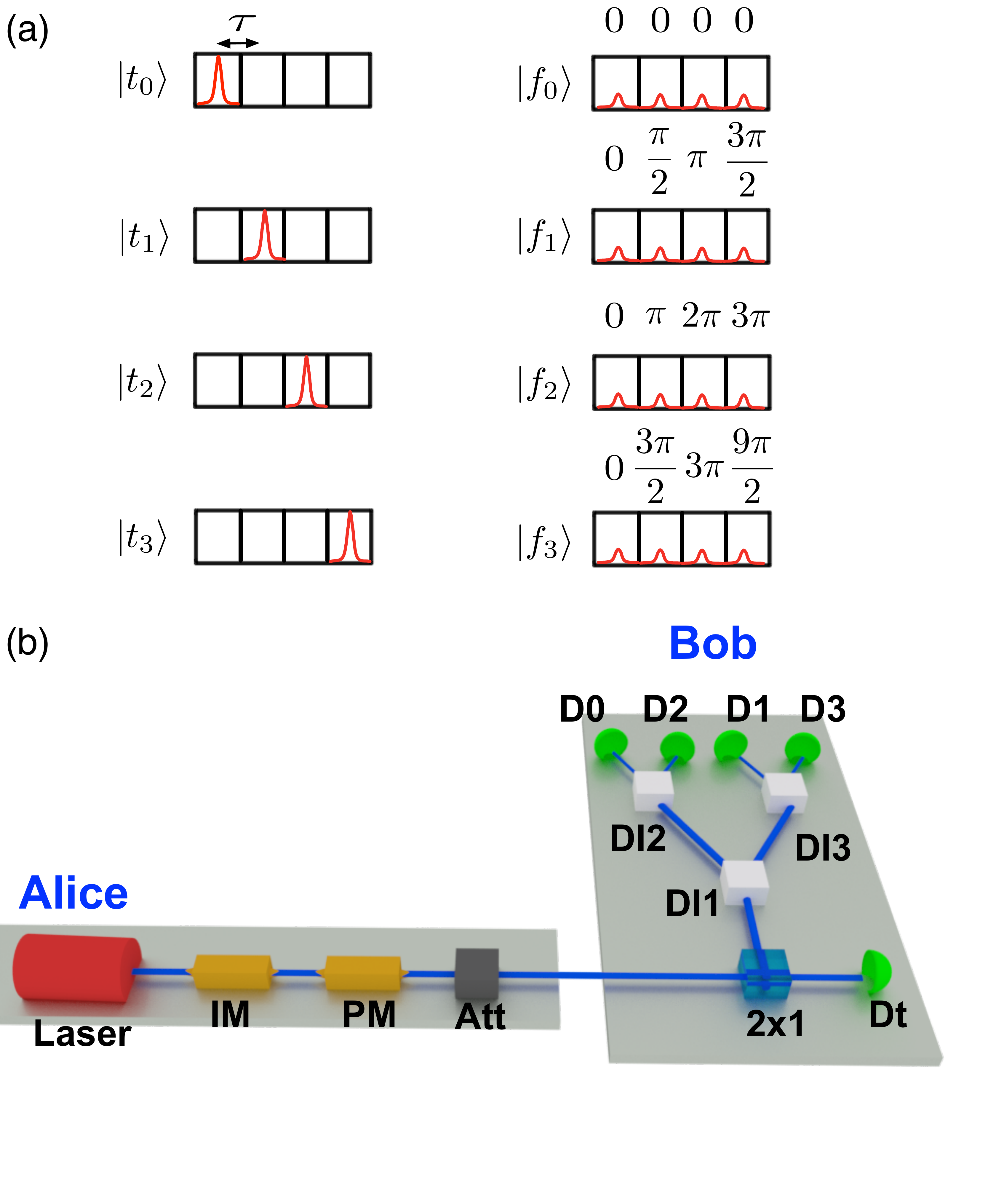}
		\caption{(a) Illustration of the four dimensional time-phase states. (b) Experimental setup of $d = 4$ time-phase QKD.}
		\label{ExperimentalSetup}
	\end{center}
\end{figure}

To demonstrate the applicability of our efficient method to a real QKD system, we consider the use of time-phase states as implemented recently in a high-rate QKD~\cite{Taimur2017}. In this experiment, the time states encode the information, and the phase basis states monitor for an eavesdropper. The $d = 4$ time-basis states are coherent-state wavepackets of duration 66~ps localized to one of the four contiguous time bins as shown in Fig.~\ref{ExperimentalSetup}(a). The time-bin width $\tau$ is set to 400~ps so that the symbol duration is 1.6~ns. The phase-basis states are given by Eq.~\ref{DFT}.

%*** Error above - using $\tau$ for time bin, but earlier defined as symbol time. ***

%Should be fixed now

%Figure~\ref{ExperimentalSetup}(b) shows a schematic of the experimental setup used to implement the protocol. The quantum states are generated by driving intensity and phase modulators (IM and PM) at a repetition rate of $r = 625$~MHz using a sequence of arbitrary patterns loaded on a field-programmable gate array (FPGA, not shown here for clarity). The states are then attenuated to the single-photon level using a variable optical attenuator (ATT) and transmitted through a quantum channel to Bob. At the receiver, Bob uses a directional coupler to direct 90\% of the quantum states for time-bin basis measurement and 10\% for the phase basis measurement. The time-basis states are measured using low timing-jitter, single-photon counting detectors (Dt) connected to a high-resolution time-to-digital converter, and the phase basis measurement scheme requires three time-delay interferometers (DIs) coupled into single-photon counting detectors (D0, D1, D2, and D3). 

Figure~\ref{ExperimentalSetup}(b) shows a schematic of the experimental setup used to implement the protocol. The quantum states are generated by modulating the intensity and phase of a continuous wave laser (1550~nm) using three electro-optic intensity modulators (IM, only one shown for Clarity) and one phase modulator (PM). All electro-optic modulators are from EOSpace. The first IM is driven with a 5~GHz sine-wave signal generator to create 66~ps-width pulse train. The second intensity modulator is used to create the time and phase states, and the third IM (not shown) is used to generate the decoy intensities. The mean photon numbers for the signal, decoy and vacuum states are set to 0.66, 0.16 and 0.002, respectively, at all channel losses except at 4~dB loss where the mean photon numbers are set to 0.45, 0.12 and 0.002. The repetition rate of the states are set to $r = 625$~MHz using a sequence of arbitrary patterns loaded on a field-programmable gate array (FPGA, not shown here for clarity). The states are then attenuated to the single-photon level using a variable optical attenuator (Att) and transmitted through a quantum channel to Bob. At the receiver, Bob uses a directional coupler to direct 90\% of the quantum states for time-bin basis measurement and 10\% for the phase basis measurement. The time-basis states are measured using low timing-jitter, single-photon counting detectors (Dt) connected to a high-resolution (50~ps) time-to-digital converter (Agilent Acqiris U1051A), and the phase basis measurement scheme requires three time-delay interferometers (DIs) coupled into superconducting nanowire single-photon detectors (SNSPDs D0, D1, D2, and D3)~\cite{Taimur2016}. The detectors used in this experiments have high detection efficiency ($>$70\%), low timing-jitter ($<$ 50 ps) and low dark count rates ($<$ 100 cps). 

\begin{figure}[htp]
	\begin{center}
		\includegraphics[width=\linewidth]{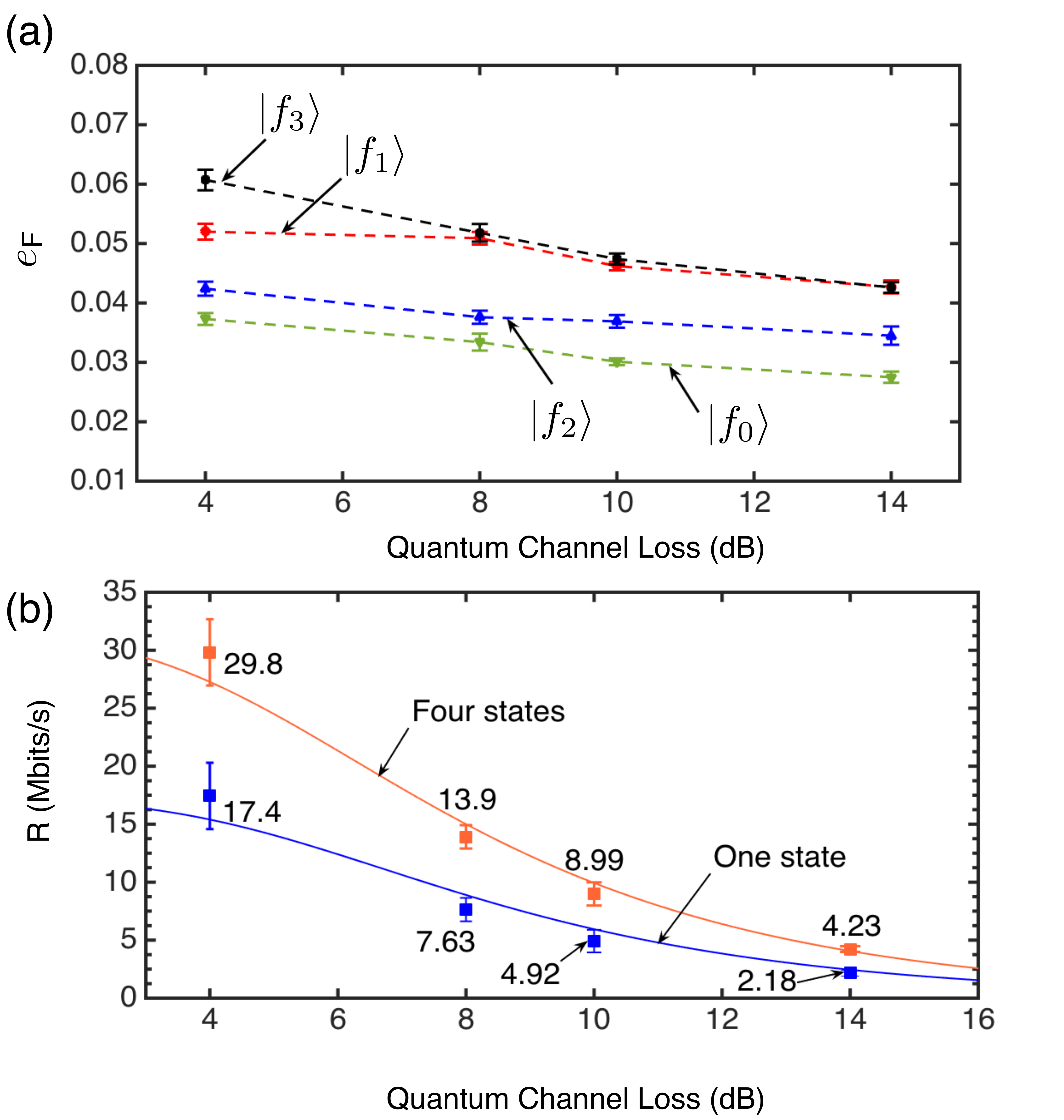}
		\caption{(a) The error rates corresponding to each of the phase basis states plotted as a function of the channel loss. (b) Asymptotic secret key rates for the cases where Alice transmits one or four phase basis states are plotted as a function of quantum channel loss. }
		\label{SKR}
	\end{center}
\end{figure}

In the experiment, all eight time and phase basis states are generated. However, the error rate in the phase basis are state-dependent, that is, some states are generated and detected more accurately than the others, mainly due to experimental challenges associated with generating the phase-basis states. Specifically, to generate $|f_1\rangle$, $|f_2\rangle$, and $|f_3\rangle$, three independent FPGA signals need to be combined at a 3$\times$1 coupler and the output signal is then used to drive the phase modulator. Each individual signal from the FPGA propagates through a different path and thus arrive at the coupler at a different time. Therefore, the combined signal used to drive the phase modulator is not perfect and results in imperfect phase values. In addition, the phase of the delay interferometers is not accurate, which can lead to state-dependent error rates as well. The error rate in phase basis as a function of the quantum channel loss is shown in Fig.~\ref{SKR}(a). 

Here, we consider the effect of reducing the number of phase-basis states on the extractable secret key rate. Suppose that Alice only transmits the state with the lowest error rate ($|f_0\rangle$ in this case) to determine the presence of an eavesdropper, which reduces the average quantum bit error rate in the phase basis. However, the bound for the phase error rate when only one state is transmitted is always higher (worse) than the bound when all states are transmitted, as shown in Fig.~\ref{Security_Proof}(b). This means that the secret key rate in the case where only one state is transmitted may be reduced even if the average quantum bit error rate is very small. 
This is illustrated in Fig.~\ref{SKR}(b), where we plot the experimentally extractable secret key rate as a function of channel loss. The orange data points represent the extractable secret key rates when all four phase-basis states are transmitted; the blue data points represent the secret key rates when only $|f_0\rangle$ is transmitted. The dashed lines represent the simulated secret key rates calculated using parameters that match the experimental conditions. As expected, it is seen that the secret key rate is smaller for the case where only state $|f_0\rangle$ is transmitted to monitor the presence of an eavesdropper. However, the reduction of the secret key at 4~dB channel loss is only $41.6\%$ compared to the case where all four phase-basis states are transmitted. In many implementations of QKD systems, the simplification of the experimental setup at the cost of a moderately lower secret key rate may be beneficial, especially for situations where the highest possible secret key rate is not the most important metric. 

\section{Simulation of Secret Key Rates}
To illustrate the advantage of the high-dimensional time-phase protocol with one monitoring basis state, here we simulate the secret key rate for $d = 2$ and $d = 4$ for two specific cases. First, we consider the case when Bob's detectors operate at 75\% detector efficiency independent of the incoming photon rate; that is, we assume that the detectors do not saturate. We also consider the case when Bob's detector saturate as the incoming photon rate exceeds a few mega-counts-per-second (Mcps), as is true for most practical systems~\cite{Taimur2017}.

We provide the details of our simulation in Appendix D. In our simulation, we use the channel model provided in Ref.~\cite{Taimur2017}, and model the detector saturation using a hyperbolic tangent function. Specifically, we model the photon detection rate as a function incoming photon rate as $a \tanh (r_x/b)$, where $r_x$ is the expected photon rate, and $a$ and $b$ are fit parameters obtained from calibration. For a single-pixel SNSPD, we find $a=6.5~\pm~0.7$ MHz and $b=8.63~\pm 1.80$ MHz. 

\begin{figure}[htp]
	\begin{center}
		\includegraphics[width=\linewidth]{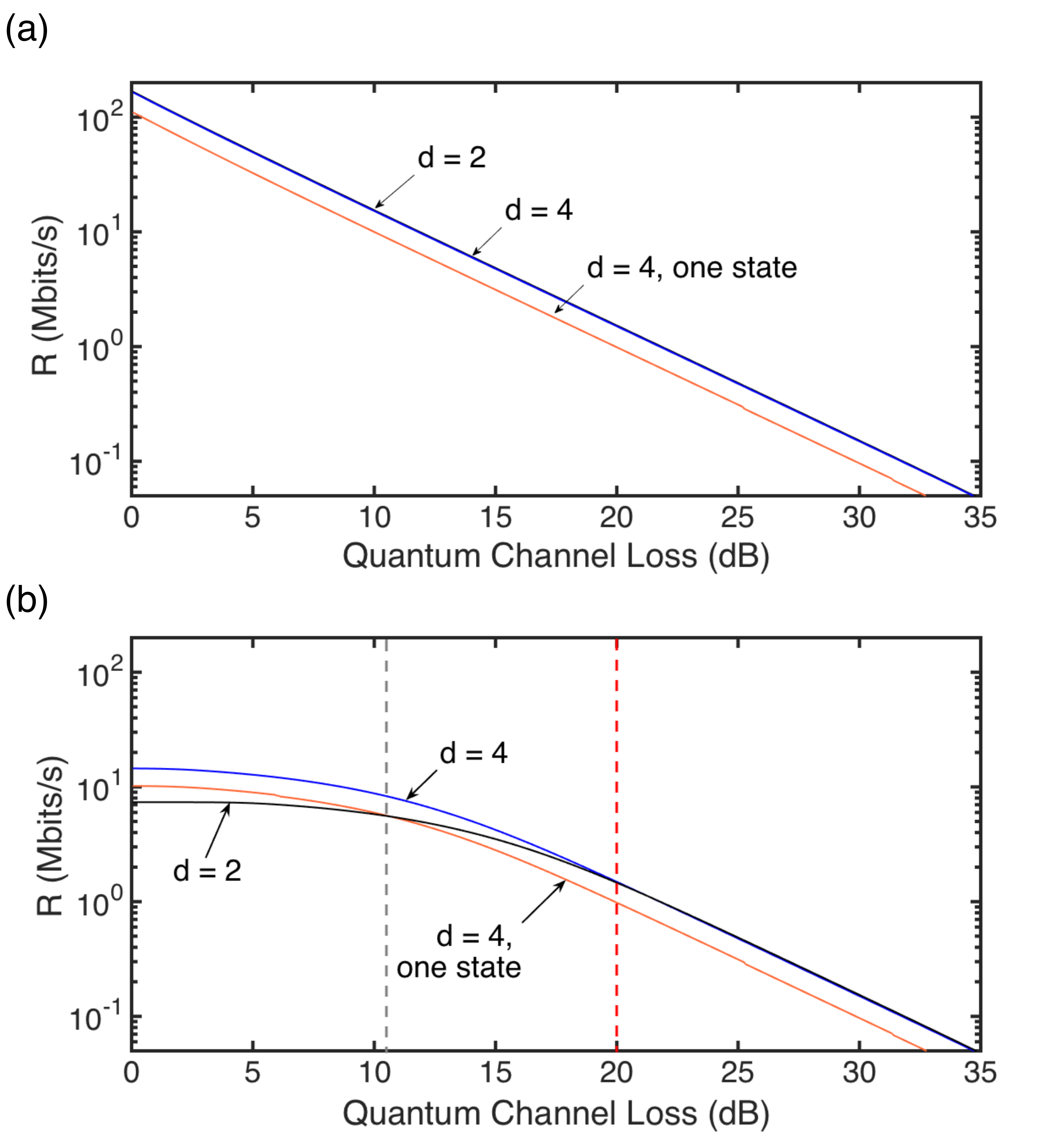}
		\caption{Dependence of secret key rate on the quantum channel loss when (a) the detectors are assumed to have a fixed efficiency of 75\% and (b) the detectors are assumed to have a rate-dependent efficiency. }
		\label{SKR_Simulated}
	\end{center}
\end{figure}

Given these two detector models, we simulate the secret key rates for $d = 2$ and $d = 4$. For $d = 4$, we consider the cases where Alice transmits all four monitoring states as well as just one monitoring state. Figure~\ref{SKR_Simulated}(a) shows the dependence of secret key rate as a function of the quantum channel loss when the detectors are assumed to be operating at 75\% efficiency, independent of the incoming photon rate. We observe that there is no difference in secret key generation rate between $d = 2$ and $d = 4$ when all monitoring basis states are transmitted. When only one monitoring basis state is transmitted, the $d = 4$ protocol has a lower secret key rate than $d = 2$ protocol. 

In Fig.~\ref{SKR_Simulated}(b), we plot the secret key rate as a function of channel loss assuming detector saturation, which is a practical problem. It is seen that when complete sets of monitoring basis states are used, a higher secret key rate can be generated using $d = 4$ than $d = 2$ upto a quantum channel loss of $\sim 20$ dB, as indicated by red dashed vertical line. Additionally, the secret key rate when only one monitoring basis is transmitted in $d = 4$ is also higher than $d = 2$ until $\sim$ 10.5 dB channel loss (gray dashed line). These results show that in the regime of detector saturation, one monitoring basis state in $d = 4$ generates higher secret key rate than $d = 2$. 

\section{Conclusion}
In conclusion, we demonstrate that it is possible to secure a high-dimensional QKD system when using less than complete set of mutually unbiased basis states. For systems with low error, we show that it is possible to use only one monitoring-basis state, which offers substantial advantages in the implementation of high-dimensional systems. Another implication of our method is that many of the current QKD systems, such as the time-bin-encoded variant of BB84~\cite{Lucamarini13}, and the coherent one-way QKD protocols~\cite{Boris2014}, can be easily upgraded to high-dimension protocols with simple modifications to the setup, thereby increasing the secure key rate of these systems.  

\section*{Acknowledgments}
We gratefully acknowledge the financial support of the ONR MURI program on Wavelength-Agile Quantum Key Distribution in a Marine Environment (Grant \# N00014-13-1-0627), and the DARPA DSO InPho program. C.C.W.L acknowledges support from National University of Singapore startup grant (No. R-263-000-C78-133/731) and CQT fellow grant (No. R-710-000-027-135).

%\section*{Supplemental Documents}
%See Supplement 1 for supporting content.
%\bigskip \noindent See \href{link}{Supplement 1} for supporting content.

%\section*{References}
% Bibliography

\section*{Appendices}
\subsection{Security Analysis Framework}
We analyze the security of the protocol by casting the problem into an optimization framework~\cite{COW12}. As discussed in the main text, the procedure requires values of all statistics known \textit{a priori} to maximize the single-photon phase-error rate, which quantifies the amount of information Eve has about Alice's measurement outcomes. 

Analysis of qudit protocols requires promoting Pauli matrices, which are used as the Hilbert-Schmidt basis for qubit-based protocols, to higher dimensions. Specifically, we require a high-dimensional basis set that satisfies two specific criteria: 1) excluding the identity matrix, all other Pauli-equivalent matrices in the high-dimensional space must be traceless; and 2) the matrices must be orthogonal. One set of matrices that satisfies these requirements are the Weyl operators~\cite{Gernot05}. For a $d$-dimensional system, the $d^2$ of basis operators are given by
\begin{align}
U_{nm} = \sum_{k=0}^{d-1}e^{\frac{2\pi i k n}{d}} |k\rangle \langle k+m|,~~~n,m = 0, 1, ..., d-1,
\end{align}
where states $|k\rangle$ and $|k+m\rangle$ are in the computational basis. %The operator $U_{01}$ $(U_{10})$ represents the matrix whose eigenstates are the temporal (phase) basis states.  

In an equivalent entanglement distillation version, Alice prepares an entangled state of the form
\begin{align}
|\phi\rangle_{AB} = \frac{1}{\sqrt{d}} \sum_{k = 0}^{d-1} |k\rangle_A |k\rangle_B,
\end{align}
where she chooses to measure either in $\{\mathsf{T}, \mathsf{F}\}$ and Bob chooses to measure in $\{\mathsf{T}, \mathsf{F}^*\}$. Note that the random basis choice of $\{\mathsf{T}, \mathsf{F}\}$ means that Alice performs the identity operation $\mathsf{I}$ for $\mathsf{T}$ and rotation $\mathcal{H}$ for $\mathsf{F}$ before sending the state to Bob. Here, $\mathcal{H}$ is an operator that performs the discrete Fourier transformation. Upon arrival of the signal states, Bob performs a unitary operation $\mathsf{I}$ for $\mathsf{T}$ and $\mathcal{H}^{-1}$ for $\mathsf{F}$.

We assume that Eve's interaction with Alice and Bob's qudits is independent and identically distributed (i.i.d) such that Eve holds an ancillary on Alice and Bob's qudits. This choice of modeling addresses so-called collective attacks. The fact that we assume Eve interacts with Alice and Bob qudits i.i.d means that the state shared among Alice, Bob and Eve can be written as $|\Psi\rangle_{ABE} = \sum_j \sqrt{\lambda_j} |\phi\rangle_{AB} |j\rangle$, such that Eve holds a purification of an ancillary characterized through the density operator
\begin{align}
\rho_E = \text{Tr}_{AB} [~|\Psi\rangle_{ABE}\langle \Psi|~]. 
\end{align}

It is well known that one can promote the security for collective attacks to coherent attacks if the quantum states are permutationally invariant using techniques like the quantum de Finetti theorem \cite{Renner05,Renner07,Renner09}. We note that the quantum states shared between Alice and Bob consists of blocks of time bins or frames, which consists of vital coherence information. These states are permutationally invariant only if they are considered as frames and not as individual time bins. 

Eve's interaction with Alice and Bob's qudits introduces an average disturbance $e_{\mathsf{X}}$, where $\mathsf{X}\in\{\mathsf{T}, \mathsf{F}\}$ represents the basis. The quantum bit error can be written as
\begin{align}
e_\mathsf{T} = \text{Tr}(\rho_{AB} E_\mathsf{T}),
\end{align}
and the phase error can be written as
\begin{align}
e_\mathsf{F} = \text{Tr}(\rho_{AB} E_\mathsf{F}),
\end{align}
where $E_\mathsf{X}$ are error operators in the $\mathsf{X}\in\{\mathsf{T}, \mathsf{F}\}$ basis. The error operator in the $\mathsf{T}$ basis and is given by,
\begin{align}
E_{\mathsf{T}} = \sum_{l\in\{0,...,d-1\}} \sum_{k\in\{0,...,d-1\}^*} |l_A, (l+k)_B\rangle \langle l_A, (l+k)_B|
\end{align}
The asterisk in the index of the sum represents the fact that all the indices should follow the general rule $l+k \neq k$. Similarly, the error operator in the $\mathsf{F}$ basis can be written as
\begin{align}
E_{\mathsf{F}} =  (\mathcal{H}_A^\dagger \otimes \mathcal{H}_B)  E_{\mathsf{T}} (\mathcal{H}_A \otimes \mathcal{H}_B^\dagger).
\end{align}
Finally, the projectors can be written as
\begin{align}
P_{l} = |l\rangle \langle l|~~~~~~~~~\text{and}~~~~~~~~~P_{\tilde{l}} = |\tilde{l} \rangle \langle \tilde{l}|
\end{align}
for both the $\mathsf{T}$ and $\mathsf{F}$ bases.

\subsection{Explicit Calculation for $d = 4$}
In this section, we go through the explicit steps for the security analysis for $d=4$ time-phase states.  The temporal states ($\mathsf{T}$-basis states) are denoted by $|t_0\rangle, |t_1\rangle, |t_2\rangle, |t_3\rangle$. The corresponding phase states ($\mathsf{F}$-basis states) are then given by
\begin{align}
|f_0\rangle &= \frac{1}{2}(|t_0\rangle + |t_1\rangle + |t_2\rangle + |t_3\rangle), \nonumber \\
|f_1\rangle &=  \frac{1}{2}(|t_0\rangle + i|t_1\rangle - |t_2\rangle - i|t_3\rangle), \nonumber \\
|f_2\rangle &=  \frac{1}{2}(|t_0\rangle - |t_1\rangle + |t_2\rangle - |t_3\rangle),\nonumber \\
|f_3\rangle &=  \frac{1}{2}(|t_0\rangle - i|t_1\rangle - |t_2\rangle + i|t_3\rangle) 
\end{align}
via discrete Fourier transformation.

The bit error operator in $\mathsf{T}$-basis is given by
\begin{align}
E_{\mathsf{T}} = \left(
\begin{array}{cccccccccccccccc}
0 & 0 & 0 & 0 & 0 & 0 & 0 & 0 & 0 & 0 & 0 & 0 & 0 & 0 & 0 & 0 \\
0 & 1 & 0 & 0 & 0 & 0 & 0 & 0 & 0 & 0 & 0 & 0 & 0 & 0 & 0 & 0 \\
0 & 0 & 1 & 0 & 0 & 0 & 0 & 0 & 0 & 0 & 0 & 0 & 0 & 0 & 0 & 0 \\
0 & 0 & 0 & 1 & 0 & 0 & 0 & 0 & 0 & 0 & 0 & 0 & 0 & 0 & 0 & 0 \\
0 & 0 & 0 & 0 & 1 & 0 & 0 & 0 & 0 & 0 & 0 & 0 & 0 & 0 & 0 & 0 \\
0 & 0 & 0 & 0 & 0 & 0 & 0 & 0 & 0 & 0 & 0 & 0 & 0 & 0 & 0 & 0 \\
0 & 0 & 0 & 0 & 0 & 0 & 1 & 0 & 0 & 0 & 0 & 0 & 0 & 0 & 0 & 0 \\
0 & 0 & 0 & 0 & 0 & 0 & 0 & 1 & 0 & 0 & 0 & 0 & 0 & 0 & 0 & 0 \\
0 & 0 & 0 & 0 & 0 & 0 & 0 & 0 & 1 & 0 & 0 & 0 & 0 & 0 & 0 & 0 \\
0 & 0 & 0 & 0 & 0 & 0 & 0 & 0 & 0 & 1 & 0 & 0 & 0 & 0 & 0 & 0 \\
0 & 0 & 0 & 0 & 0 & 0 & 0 & 0 & 0 & 0 & 0 & 0 & 0 & 0 & 0 & 0 \\
0 & 0 & 0 & 0 & 0 & 0 & 0 & 0 & 0 & 0 & 0 & 1 & 0 & 0 & 0 & 0 \\
0 & 0 & 0 & 0 & 0 & 0 & 0 & 0 & 0 & 0 & 0 & 0 & 1 & 0 & 0 & 0 \\
0 & 0 & 0 & 0 & 0 & 0 & 0 & 0 & 0 & 0 & 0 & 0 & 0 & 1 & 0 & 0 \\
0 & 0 & 0 & 0 & 0 & 0 & 0 & 0 & 0 & 0 & 0 & 0 & 0 & 0 & 1 & 0 \\
0 & 0 & 0 & 0 & 0 & 0 & 0 & 0 & 0 & 0 & 0 & 0 & 0 & 0 & 0 & 0 \\
\end{array}
\right),
\end{align}
and the Fourier-transform matrix for $d=4$ is given by
\begin{align}
\mathcal{H} =  \frac{1}{\sqrt{4}}\left(\begin{array}{cccc}
1 & 1 & 1 & 1 \\
1 & e^{\frac{1}{2} [-(\pi  i)]1} & e^{\frac{1}{2} [-(\pi  i)] 2} & e^{\frac{1}{2} [-(\pi  i)] 3} \\
1 & e^{\frac{1}{2} [-(\pi  i)] 2} & e^{\frac{1}{2} [-(\pi  i)] 4} & e^{\frac{1}{2} [-(\pi  i)] 6} \\
1 & e^{\frac{1}{2} [-(\pi  i)] 3} & e^{\frac{1}{2} [-(\pi  i)] 6} & e^{\frac{1}{2} [-(\pi  i)] 9} \\
\end{array}
\right).
\end{align}
Therefore, the error operator for the phase basis is given by $E_{\mathsf{F}} =  (\mathcal{H}_A^\dagger \otimes \mathcal{H}_B)  E_{\mathsf{T}} (\mathcal{H}_A \otimes \mathcal{H}_B^\dagger)$. All the joint probabilities where Alice sends states in one basis and Bob measures in the other is equal \texttt{1/16}. All the joint probabilities for partial measurements, where Alice sends states in the $\mathsf{T}$-basis and Bob measures in the $\mathsf{F}$-basis are equal to \texttt{1/4(1-$e_\mathsf{T}$)} for correct measurements, and equal to \texttt{1/12$e_\mathsf{T}$} for incorrect measurements.  

\subsection{Pseudo-Code for SDP}
The algorithm for the SDP code is as follows: \\
\begin{center}
	\texttt{
		maximize Tr($E_\mathsf{F} \rho_{AB}$) such that, \\
		Tr($\rho_{AB}$) = 1\\
		$\rho_{AB} \geq 0$ \\
        \vspace{10pt}
		\%QBER is the expected experimental error \\
         \vspace{10pt}
		Tr($E_\mathsf{T} \rho_{AB}$) = $e_\mathsf{T}$ \\
         \vspace{10pt}
		\%All the joint operators where Alice sends F and Bob measures in T \\
        \%The indices i, j = 1:4\\
        \vspace{10pt}
		Tr$(|f_i\rangle\langle f_i|\otimes|t_j\rangle \langle t_j| \rho_{AB}) = \frac{1}{16}$ \\
         \vspace{10pt}
  \vspace{10pt}
		\%All the joint operators where Alice sends T and Bob measures in F \\
        \%The indices i, j = 1:4\\
        \vspace{10pt}
		Tr$(|t_i\rangle\langle t_i|\otimes|f_j\rangle \langle f_j| \rho_{AB}) = \frac{1}{16}$ \\
         \vspace{10pt}
		\%All the partial measurements where Alice sends F and Bob measures in F \\
         \vspace{10pt}
		~~~~Tr$(|f_0\rangle\langle f_0|\otimes|f_0\rangle \langle f_0| \rho_{AB}) = 1/4\times(1-e_\mathsf{T})$ \\	
		Tr$(|f_0\rangle\langle f_0|\otimes|f_1\rangle \langle f_1| \rho_{AB}) = 1/12\times e_\mathsf{T}$ \\	
		Tr$(|f_0\rangle\langle f_0|\otimes|f_2\rangle \langle f_2| \rho_{AB}) = 1/12\times e_\mathsf{T}$ \\	
		Tr$(|f_0\rangle\langle f_0|\otimes|f_3\rangle \langle f_3| \rho_{AB}) = 1/12\times e_\mathsf{T}$ \\	
		Tr$(|f_1\rangle\langle f_1|\otimes|f_0\rangle\langle f_0| \rho_{AB}) = 1/12\times  e_\mathsf{T}$ \\	
		~~~~Tr$(|f_1\rangle\langle f_1|\otimes|f_1\rangle\langle f_1| \rho_{AB}) = 1/4\times (1-e_\mathsf{T})$ \\	
		Tr$(|f_1\rangle\langle f_1|\otimes|f_2\rangle\langle f_2| \rho_{AB}) = 1/12\times e_\mathsf{T}$ \\	
		Tr$(|f_1\rangle\langle f_1|\otimes|f_3\rangle\langle f_3| \rho_{AB}) = 1/12\times e_\mathsf{T}$ \\	 
		Tr$(|f_2\rangle\langle f_2|\otimes|f_0\rangle\langle f_0|\rho_{AB}) = 1/12\times e_\mathsf{T}$ \\	
		Tr$(|f_2\rangle\langle f_2|\otimes|f_1\rangle\langle f_1| \rho_{AB}) = 1/12\times  e_\mathsf{T}$ \\	
		~~~~Tr$(|f_2\rangle\langle f_2|\otimes|f_2\rangle\langle f_2| \rho_{AB}) = 1/4\times  (1-e_\mathsf{T})$ \\
		Tr$(|f_2\rangle\langle f_2|\otimes|f_3\rangle\langle f_3| \rho_{AB}) = 1/12\times e_\mathsf{T}$ \\
		Tr$(|f_3\rangle\langle f_3|\otimes|f_0\rangle\langle f_0| \rho_{AB}) = 1/12\times e_\mathsf{T}$ \\	
		Tr$(|f_3\rangle\langle f_3|\otimes|f_1\rangle\langle f_1| \rho_{AB}) = 1/12\times e_\mathsf{T}$ \\	
		Tr$(|f_3\rangle\langle f_3|\otimes|f_2\rangle\langle f_2| \rho_{AB}) = 1/12\times e_\mathsf{T}$ \\	
		~~~~Tr$(|f_3\rangle\langle f_3|\otimes|f_3\rangle\langle f_3| \rho_{AB}) = 1/4\times (1-e_\mathsf{T})$
	}	
\end{center}
The last sixteen lines show explicitly the partial measurements of only a subset of the $\mathsf{F}$-basis states. 

\subsection{Channel Model}
We adopt the channel model described in Ref.~\cite{Taimur2017}. Specifically, we model the probability of Alice transmitting a state and Bob receiving it as
\begin{align}
R_\mathsf{X, k} = [1-\exp(-\eta_{det} \eta_{ch} k)]+ P_d/d,
\end{align}
where $\eta_{ch}$ is the transmission of the quantum channel, $\eta_{det}$ is the efficiency of the single-photon detector, $k \in \{\mu, \nu, \omega\}$ is the mean photon number, $\mathsf{X} \in \{\mathsf{T}, \mathsf{F}\}$ is the basis, and $P_d$ is the probability of detecting an event due to detector dark counts. The overall gain can be expressed as
\begin{align}
R_\mathsf{X} = p_{\mu} R_\mathsf{X, \mu} + p_{\nu} R_\mathsf{X, \nu} + p_{\omega} R_\mathsf{X, \omega}.
\end{align}
Similarly, the probability of Bob receiving an incorrect state can be written as
\begin{align}
E_\mathsf{X, k} = e_{d}[1-\exp(-\eta_{det} \eta_{ch} k)]+ (d-1) P_d/d,
\end{align}
where $e_d$ is the intrinsic error rate. The overall error rate can be written as
\begin{align}
E_\mathsf{X} = p_{\mu} E_\mathsf{X, \mu} + p_{\nu} E_\mathsf{X, \nu} + p_{\omega} E_\mathsf{X, \omega}.
\end{align}
The secret key fraction is then simulated by calculating the bounds for $R_{\mathsf{T,1}}$, $e^U_{\mathsf{F}}$, $\Delta_{\mathsf{Leak}}:= E_{\mathsf{T}}/R_{\mathsf{T}}$ and then calculating $K$ in Eq.~\ref{SKF_Time_Phase}. The secret key rate can then be determined by calculating $r K$, where $r$ is the state preparation rate. 

In our simulations shown in Sec. IV, we use $P_d = 10^{-7}$, $e_d= 0.005\times d$, $r= 2500/d$~MHz, $P_{\mu}= 0.8,~P_{\nu}= 0.1$ and $P_{\nu}= 0.1$. The probability of transmitting time and phase basis states are set to $P_{\mathsf{T}} = 0.9$ and $P_{\mathsf{F}} = 0.1$ to maximize the secret key rate~\cite{Lucamarini13}. We optimize the mean photon numbers at every channel loss using the Matlab function Fmincon. We set the mean photon numbers between 0.05 and 0.97 under the constraint $\nu + \omega < \mu$ and $0\leq \omega \leq \nu$. For the case where we assume that the detector efficiencies are independent of the incoming photon rate, we set $\eta_{det} = 0.75$. When we take the detector saturation into account, we assume that the dependence of the detection rate as a function of the expected rate can be modeled as $a\tanh({r_x/b})$ as discussed in the main text. 

\bibliography{sample}
\end{document}